\documentstyle[12pt, openbib]{article}

\title{Scattering  of positronium by H, He, Ne, and Ar}

\author{P. K. Biswas and Sadhan K. Adhikari\\
Instituto de F\'{\i}sica Te\'orica, Universidade Estadual
Paulista,\\
01.405-900 S\~ao Paulo, S\~ao Paulo, Brazil\\}

\date{\today}
\begin{document}
\maketitle
\begin{abstract}

The low-energy scattering of ortho positronium (Ps) by H, He, Ne, and Ar
atoms has been investigated in the coupled-channel framework by using a
recently proposed
time-reversal-symmetric nonlocal electron-exchange model potential with a
single parameter $C$. For H and He we use a three-Ps-state
coupled-channel
model and for Ar and Ne we use a static-exchange model. The sensitivity of
the results is  studied 
with respect to the parameter $C$.
Present low-energy cross sections for He, Ne and Ar are in good agreement
with experiment.

\end{abstract} 

\newpage


Low-energy collision of ortho positronium (Ps) 
atom with neutral gas atoms and
molecules 
is of  interest in both physics and chemistry.  Recently, there have been
precise
measurements of low-energy ortho-Ps scattering by H$_2$, N$_2$, He, Ne, Ar,
C$_4$H$_{10}$, and C$_5$H$_{12}$ \cite{1a,1b,2}.  Due to internal charge
and mass
symmetry, Ps atom yields zero   elastic and even-parity
transition 
potentials  in the direct channel.
Ps scattering  is 
 dominated  mainly 
by 
exchange correlation  at low
energies. 
If $N$ basis states are included for
both Ps and target in  a coupled-channel formalism, the number of channels
grow as 
$N^2$. This complicates the tractability of the Ps scattering process in a 
coupled channel scheme compared to the electron scattering. 
The dominance of the short-range interaction in Ps scattering  causes
serious trouble
towards the
convergence of any coupled-channel formalism with
a truncated  basis \cite{ad1,ad2}. 
The use of 22 coupled Ps pseudostates for Ps-H system in the R-matrix
approach has indicated convergence difficulties  \cite{ad2,wl} in 
 Ps-H binding and resonance energies. 

To find a solution to the
nonconvergence  problem, we have suggested a
nonlocal electron-exchange model potential \cite{ad1,ad2}
with a single parameter $C$ and 
demonstrated its  effectiveness 
by  performing   quantum coupled-channel
calculations using the ab initio framework of close coupling approximation. 
Two versions of this potential were
suggested: one is  time-reversal-symmetric and the other nonsymmetric.
The nonsymmetric potential has been applied for the study of Ps scattering 
by H \cite{ad4}, He \cite{ad1} and H$_2$ \cite{ad3} using a three-Ps-state
coupled-channel model. For He and H$_2$ 
these studies yielded total cross sections in good agreement with
experiments
\cite{1a,1b,2} in addition to producing the correct pick-off quenching
rate
for Ps-He at  low energies \cite{ad5}. 
Higher  Ps-excitation and ionization cross sections
were also calculated in these cases to reproduce the total  cross sections
at medium and high energies. Previous theoretical studies \cite{4a,4}
on Ps-He
produced results in strong disagreement \cite{ad1} with experiment
\cite{1a,1b,2}.

In a subsequent application of this model potential to Ps scattering by H 
\cite{ad2}, it was found that the symmetric form leads to by far superior 
result than the nonsymmetric form. The symmetric form was able to 
reproduce accurate
variational results very precisely for singlet Ps-H binding and resonance
energies; the nonsymmetric form failed to yield such results.

In view of this we reinvestigate the problem of low-energy Ps scattering 
by He \cite{ad1} using the symmetric exchange potential
employing the
three-Ps-state model used before. We study how
the low-energy cross sections for Ps-H and Ps-He change with the variation
of the parameter
$C$ of the potential. 
Then we  also apply this exchange
potential to the study of low-energy Ps scattering by Ne and Ar
using a static-exchange model. 
The present calculation accounts for the measured low-energy cross
sections \cite{1a,1b,2} satisfactorily for He, Ne, and Ar.

The total  wave function of the Ps-target system is
expanded in terms of
the 
Ps eigenstates as 
\begin{eqnarray}
\Psi^\pm({\bf r_1,...,r_N;r_{N+1},x
})&=&
\sum_\nu\biggr[ F_\nu({\bf \rho_{N+1}})\chi_\nu
({\bf t_{N+1}})\phi_0({\bf r_1,...,r_j,...,r_N}) \nonumber \\ &+&
\sum_{j=1}^N (-1)^{S_{N+1},j}F_\nu({\bf \rho_{j}})\chi_\nu
({\bf t_{j}})\phi_0({\bf r_1,...,r_{N+1},...,r_N})\biggr] ,
\end{eqnarray}
where antisymmetrization   
 with respect to Ps- and target-electron coordinates has been made. 
Here ${\bf \rho_i= (x+r_i)}/2$,  ${\bf t_i= (x-r_i)}$, where ${\bf r_i},
i=1,...,N$
denote
the target electron coordinates,  and ${\bf r_{N+1}}$ and ${\bf x}$ are 
the electron and positron coordinates of Ps;
$\phi_0$ and $\chi_\nu$ denote the target and Ps wave functions, and 
$F_\nu$ is the continuum orbital of Ps with respect to the target. 
$S_{N+1,j}$ is the total spin of the Ps electron ($N+1$) and target
electron ($j$) undergoing exchange and can have values of 1 or 0. 
The spin of the positron is conserved in this process and the exchange
profile of the Ps-target system is analogous to the corresponding
electron-target system.
Projecting the
resultant Schr\"odinger equation on the Ps eigenstates and 
 averaging over   spin states, the resulting 
momentum-space Lippmann-Schwinger scattering equation for a particular 
total electronic spin state $S$ can, in general, be
written as \cite{4,gh}
\begin{eqnarray}\label{aa} &{}&f^{S}_{\nu ' \nu}({\bf k',k})={\cal
B}^{S}_{\nu '\nu}({\bf k ',k}) \nonumber \\ &-&\frac{1}{2\pi^2}\sum_{\nu
''} \int \makebox{d}{\bf k ''}\frac {{\cal B}_{\nu' \nu ''}^S({\bf
k',k''}) f_{\nu '' \nu}^S({\bf k'',k}) } {
k_{\nu ''}^2/4
-k''^2/4+\makebox{i}0}, \end{eqnarray} 
where 
$f_{\nu'\nu}^S$ is the scattering amplitude, and ${\cal B}_{\nu '\nu}$ is
the corresponding Born potential, 
$\nu$ and $\nu '$ denote initial and final Ps states, 
$k_{\nu ''} = \sqrt{2m({\cal E}-\epsilon '')/\hbar^2} $ is the on-shell
relative
momentum of Ps in the channel $\nu ''$
with  $\epsilon{''} $
 the total binding energy of
the intermediate Ps and target states, and ${\cal E}$  the total energy
of the  
Ps-target system and $m$ the reduced mass. 
Here
\begin{equation}\label{xx}
 {\cal B}^{S}_{\nu' \nu}({\bf k',k}) = B^{D}_{\nu' \nu}({\bf
k',k})+\sum_{j=1}^N(-1)^{S_{N+1,j}} B^{E_j}_{\nu' \nu}({\bf k',k});
\end{equation}
where $B^D$ is the direct
Born potential and $B^{E_j}$ is the model  potential for exchange between  
the
Ps electron (denoted by $N+1$) with the $j$th target electron. 
For Ps-H scattering
both $S\equiv S_{2,1}$ = 0, 1   will 
contribute  \cite{gh}; the corresponding potentials and 
amplitudes are usually denoted by ${\cal B}^{\pm}_{\nu' \nu}$
and ${f}^{\pm}_{\nu' \nu}$, respectively \cite{ad2}.
For Ps scattering from He, Ne, and Ar etc. there will be only one
scattering equation (\ref{aa}) corresponding to total electronic spin 
$S=1/2$. 
For these targets with doubly occupied spatial orbitals, in
the sum
over $j$ 
in Eq. (\ref{xx}) only half of the occupied target electrons in
each sub-shell will  
contribute 
 when the target is frozen to its
ground state \cite{rt}. 
Consequently,  only   $S_{N+1,j}$ = 1 ($f^-$ and  ${\cal B}^-$) will
contribute to target-elastic scattering \cite{4} for targets with doubly
occupied
spatial orbitals.

For Ps-H scattering, the differential cross section is given by
$d\sigma/d\Omega = [|f^+|^2+3|f^-|^2]/4$. 
For target-elastic Ps scattering
by He, Ne, and Ar, 
the differential cross section is given by
$d\sigma/d\Omega = |f^-|^2$. 
In all Ps-scattering
 the direct potential, $B^D$,  is  zero for 
elastic and all even-parity-state transitions of Ps.  Thus the
nonorthogonal exchange kernel   alone dominates
the solution of Eq.  (\ref{aa}) and this dominance 
 is possibly responsible for convergence difficulties to conventional
approaches based on Eq. (\ref{aa}).

The present exchange model was derived using Slater-type orbital for the
H-atom, so that a generalization from a H-target to a complex target
represented by a Hatree-Fock wave function becomes straight-forward.  For
Ps scattering from a H orbital, the model exchange
potential between the Ps-electron ($ {\bf r_2}$) and the orbital electron
($ {\bf r_1}$)  was derived from the $1/r_{12}$ term and is given
by \cite{ad1,ad2}: 
\begin{eqnarray}\label{6}
 B^E_{\mu '\nu'\mu \nu}& =&\frac{4(-1)^{l+l'}}{<D>}
\int \phi^*_{\mu'}({\bf r}_2)\exp (i {\bf Q. r}_2)\phi_\mu ({\bf r}_2)
\makebox{d}{\bf r}_2\nonumber \\ &\times&
\int \chi_{\nu'} ^*({\bf  t }_2)\exp (i{\bf Q}.{\bf  t }_2/
2)\chi_\nu ({\bf  t }_2 ) \makebox{d}{\bf  t }_2,
\end{eqnarray}
where $l$ and $l'$ are angular momenta of the initial ($\chi_\nu$) and
final ($\chi_{\nu '}$) 
Ps
states, $\phi_\mu$ and $\phi_{\mu '}$ are initial and final H states, 
and 
${\bf Q= k_i-k_f}$. Here ${\bf k_i}$ and ${\bf k_f}$ are initial and final 
Ps momenta, respectively. In Eq. (\ref{6})
 the symmetric form of the averaged quantity $<D>$ is \cite{ad1,ad2}
\begin{equation}\label{3}
<D>=\frac{k_i^2+k_f^2}{8}+C^2\left[\frac{\alpha_\mu^2+\alpha_{\mu'}^2}{2}
+\frac{\beta_\nu^2+\beta_{\nu'}^2}{2}\right],
\end{equation}
where $\alpha_{\mu'}^2/2$ and
$\beta_\nu^2$
are the binding energies of the final target and initial Ps  states,
respectively, and  $C$ is the only parameter of the potential. Normally,
the parameter $C$ is taken to be unity \cite{ad1,ad2} which leads to
reasonably good numerical results. However, it can be varied slightly from
unity to obtain a precise  fit of a low-energy scattering observable
(experimental or variational), as have been done in some applications of
model potentials \cite{mo1,mo2}. A variation of $C$ from unity leads to a 
variation of the binding energy parameters ($\alpha^2, \beta^2$ etc.)
used as average values for square of  momentum \cite{ad1} in the
expression for
$\langle D \rangle$ of Eq. (\ref{3}). This, in turn, tunes the strength of
the exchange potential (\ref{6}) at low energies. At high energies this
model potential is insensitive to this parametrization  and leads to the
well-known Born-Oppenheimer form of exchange \cite{bo}. We have
turned this flexibility to good advantage  by obtaining precise
agreement with low-energy results   of Ps scattering  by  H, He, Ne,
and Ar, as we shall
see in the following.

For a complex target the  space part of the  HF wave function
\cite{9} is given by $\Psi({\bf r}_1,{\bf r}_2,...,{\bf
r}_j,\-...,{\bf r}_N)\- = {\cal A}[{ 
\phi}_1({\bf r}_1){ \phi}_2({\bf r}_2)...{ \phi}_j({\bf r}_j)...
{ \phi}_N({\bf r}_N)],
$ where ${\cal A}$ is the antisymmetrization operator.  
The position vectors
of the electrons  are ${\bf r} _i, i=1,2,...,N$ and ${ \phi_j}$'s have
the form:
${ \phi}_j({\bf r_j})=\sum_\kappa a_{\kappa j} { \phi}_{\kappa j}({\bf
r_j})$. The orbital ${ \phi}_{\kappa j}({\bf
r_j})$ is a Slater-type orbital. 
Considering proper
antisymmetrization with respect to Ps and target electrons, 
the final model exchange potential
obtained from (\ref{6})  is given by \cite{ad1}
\begin{eqnarray}\label{7}
B^{E_j}_{\nu '\nu}&=& \sum_{\kappa \kappa'} 
\frac{4a_{\kappa j} a_{\kappa ' j} (-1)^{l+l'}}{<D_{\kappa \kappa
'}>}\nonumber \\
&\times&
\int \phi_{\kappa ' j}^*({\bf r}_j) \exp (i {\bf Q. r}_j)\phi_{\kappa
j}({\bf
r}_j) 
\makebox{d}{\bf
r}_j\nonumber \\ &\times& \int \chi_{\nu`}^*({\bf  t }_j)\exp (i{\bf Q}.{\bf
t}_j/2)\chi_\nu ({\bf  t }_j) \makebox{d}{\bf  t }_j.
\end{eqnarray}
with $$<D_{\kappa \kappa '}> =
(k_i^2+k_f^2)/8+C^2[(\alpha^2_{\kappa
j}+\alpha^2_{\kappa ' j })/2+(\beta_\nu^2+\beta_{\nu`}^2)/2],$$
where $\alpha_{\kappa j }$ is the energy parameter corresponding to the
orbital 
$\phi_{\kappa  j }({\bf r_j})$ \cite{9}.

We use exact  wave functions for H and Ps,  HF wave functions  for He, Ne,
and Ar \cite{9}. After a partial-wave projection, the one-dimensional
scattering equations are solved by the method of matrix inversion.
 
First we study the effect of the variation of the parameter $C$ of the 
exchange potential in the
 Ps-H system 
using  a three-Ps-state model  with Ps(1s,2s,2p) states \cite{ad4}.
We start our discussion with the singlet S-wave resonance. In
Fig. 1 we plot the S-wave phase shift at different energies which
illustrate 
the resonance pattern obtained with different values of $C$.  The
resonance position shifts monotonically towards lower
energies with decreasing of $C$ from unity.  We have shown it in steps
where the value of $C$ is varied from unity to 0.785. The resonance
position matches with the accurate prediction of 4.01 eV \cite{ho1} for
$C=0.785$. In
Fig. 2, we plot $k \cot \delta$ versus $k^ 2$ for the corresponding
low-energy S-wave phase shifts $\delta$.  Figure 2 demonstrates how the 
improvement in the resonance position simultaneously improves the Ps-H
binding energy. For $C=1.0$ the resonance and binding energies are 4.715
eV and 0.165 eV, respectively;  for $C=0.9$ the corresponding energies are
4.470 eV and 0.445 eV, respectively.  At $C=0.785$, while the  resonance
position
is correctly fitted to 4.01 eV (Fig. 1), we obtain an approximate
binding energy of 1.02 eV from a linear extrapolation as in Fig. 2, and
0.99 eV with more precise fitting considering second order corrections,
compared to the accurate prediction of 1.0598 $-$ 1.067 eV \cite{fs}.
This behavior of the low-energy phase
shifts, which
yields simultaneously the Ps-H resonance  and binding energies,
indicates that the use of the model potential (\ref{6})  in a
coupled-channel scheme can lead to a good description of Ps-H
scattering.

We exhibit in Figs. 3 (a), (b), and (c)  the present elastic cross
sections, for Ps scattering by He, (three-Ps-state \cite{ad1}) Ne and Ar
(static-exchange), respectively, for 
$C=1$, 0.85, and 0.785. 
The value $C=0.785$ yielded the good agreement in
the case of Ps-H.  
For the  closed-shell atoms, 
a typical $C$ close to 0.85 works well for the 3-Ps-state model  in
Ps-He  and for the static-exchange model in Ps-Ar and
Ps-Ne.
Although the  present cross sections
differ  from other theoretical \cite{4a,4}
and experimental \cite{hy} works  
at
low energies for Ps-He (See, Fig. 6 of Ref. \cite{ad1}),
they 
agree well with the recent  measurements  
of Skalsey et al.  \cite{2} and unpublished work of G. Peach as
quoted in Ref. \cite{1b}.

\vskip 0.5cm

{Table I: Low-energy S-wave phase shifts in radians for Ps-He, Ps-Ne,
and Ps-Ar   for different $k$ in au. The entries for $k=0$ correspond
to the scattering lengths in units of $a_0$, incident positronium energy
$E=6.8 k^2$
eV.}
\vskip
.2cm

{\begin{center}{\begin{tabular} {|c|c|c|c|c|}
\hline
$k$  & Ar(SE)   & Ne(SE)  & He(SE) & He(3st)\\
 \hline
0.0 & 1.65    & 1.41     &  1.03       &   0.90\\
0.1 & $-$0.164  & $-$0.141 & $-$0.103& $-$0.088\\
0.2 & $ -$0.319 &$-$0.277  & $-$0.202& $-$0.172\\
0.3 & $-$0.457  &$-$0.404  &$-$0.294 & $-$0.249\\
0.4 & $-$0.572  &$-$0.518  &$-$0.375 & $-$0.315\\
0.5 & $-$0.656  &$-$0.615  & $-$0.444& $-$0.368\\
0.6 & $-$0.706  &$-$0.694  &$-$0.500 & $-$0.408\\
0.7 & $-$0.720  &$-$0.754  &$-$0.541 & $-$0.433\\
0.8 & $-$0.699  &$-$0.792  &$-$0.569& $-$0.445\\
\hline
\end{tabular}
}\end{center}}

\vskip 0.5cm
{Table II: Low-energy P-wave phase shifts in radians for Ps-He, Ps-Ne,
and Ps-Ar   for different $k$ in au. The incident positronium energy
$E=6.8 k^2$
eV. The numbers in parenthesis denote powers of ten. }
\vskip
.2cm
{\begin{center}{
\begin{tabular} {|c|c|c|c|c|}
\hline
$k$  & Ar(SE)   & Ne(SE)  & He(SE) & He(3st)\\
 \hline
0.1 & $-$2.71($-$3)  &$-$1.63($-$3) & $-$8.19($-$4) &$-6.24(-4)$ \\
0.2 & $ -$1.94($-$2) &$-$1.20($-$2) & $-$6.11($-$3)  & $-4.63(-3)$\\
0.3 & $-$5.51($-$2)  &$-$3.55($-$2) &$-$1.85($-$2) & $-1.38(-2)$\\
0.4 & $-$1.06($-$1)  &$-$7.17($-$2) &$-$3.81($-$2)   & $-2.80(-2)$\\
0.5 & $-$1.65($-$1)  &$-$1.17($-$1) & $-$6.33($-$2)   & $-4.53(-2)$\\
0.6 & $-$2.23($-$1)  &$-$1.66($-$1) &$-$9.15($-$2)  & $-6.26(-2)$\\
0.7 & $-$2.75($-$1)  &$-$2.15($-$1) & $-$1.20($-$1) & $-7.63(-2)$\\
0.8 & $-$3.17($-$1)  &$-$2.60($-$1) &$-$1.47($-$1)   & $-7.89(-2)$\\
\hline
\end{tabular}   }\end{center}}

\vskip 0.5cm

In Table I we present the S-wave phase shifts for Ps-Ar (static exchange
model denoted SE),
Ps-Ne (SE), and Ps-He (SE and three-Ps-state models)
scattering 
for $C=0.85$. In Table
II we present the same for the P wave. The magnitude of the 
scattering lengths, low-energy
cross sections and phase shifts (well below Ps-excitation threshold) 
  increase monotonically 
as we move from He to Ne and from Ne to Ar. As the effective potential
for elastic scattering in
these cases is repulsive in nature, this indicates an increase in
repulsion from He to Ne and from Ne to Ar.

We find from Figs. 3 that 
the energy-dependences of the elastic cross sections are similar for all
the closed-shell atoms studied here. The cross section has a monotonic
slow decrease
with increasing energy. This trend is consistently found in all previous
theoretical calculations in Ps-He. Also, this is expected as the
underlying effective potential for elastic scattering is repulsive in
nature.

In conclusion, we have reinvestigated the problem of low-energy elastic Ps
scattering by H, and He (three-Ps-state) using a symmetric nonlocal
electron-exchange potential with a parameter $C$. We further apply this
potential to Ps scattering by Ne, and Ar (static-exchange).  Although the
value $C=1$ was originally suggested, a slightly lower value $C\approx 0.8
$ leads to good agreement with accurate experiment \cite{2} and accurate
calculations \cite{ho1,fs} in the present cases.  Although, a
non-symmetric form of the model potential provides a fairly good account
of the cross section \cite{ad4}, we have found that the symmetric form is
able to
provide a more precise description of scattering. 
The Ref. \cite{ad1}
we
have demonstrated the effectiveness of the present exchange potential in
electron-impact scattering.  Simplicity of the present exchange potential
and the reliability of the present results calculated with it from a two-
(Ps-H) to a 19-electron (Ps-Ar) system reveals the effectiveness of the
exchange model and warrants further study with it.

We thank Prof. B. H. Bransden for his helpful and encouraging comments.
The work is supported in part by the CNPq and FAPESP of Brazil.



\vskip 0.5cm
Figure Caption

1. S-wave singlet Ps-H phase shifts in radian showing the
variation of the resonance position with the variation
of $C$ in (\ref{3}) using present three-Ps-state model.

2. $k \cot \delta $ and $ik$ versus $ k^2$ plot showing the change in Ps-H
binding energy with the variation in $C$ as in figure 1 (Energy = $6.8
k^2$ eV). The crossing of the $k \cot \delta $ and $ik$ curves give the
energy of the bound state. 

 3. Cross section of Ps scattering by  (a) He, (three-Ps-state model)
(b) Ne, and
(c) Ar (static-exchange model) for different $C$: 
   $C=1$ (full line), $C=0.85$ (dashed-dotted) line,
$C=0.785$ 
(dashed line), experiment (box, Ref. \cite{2}).

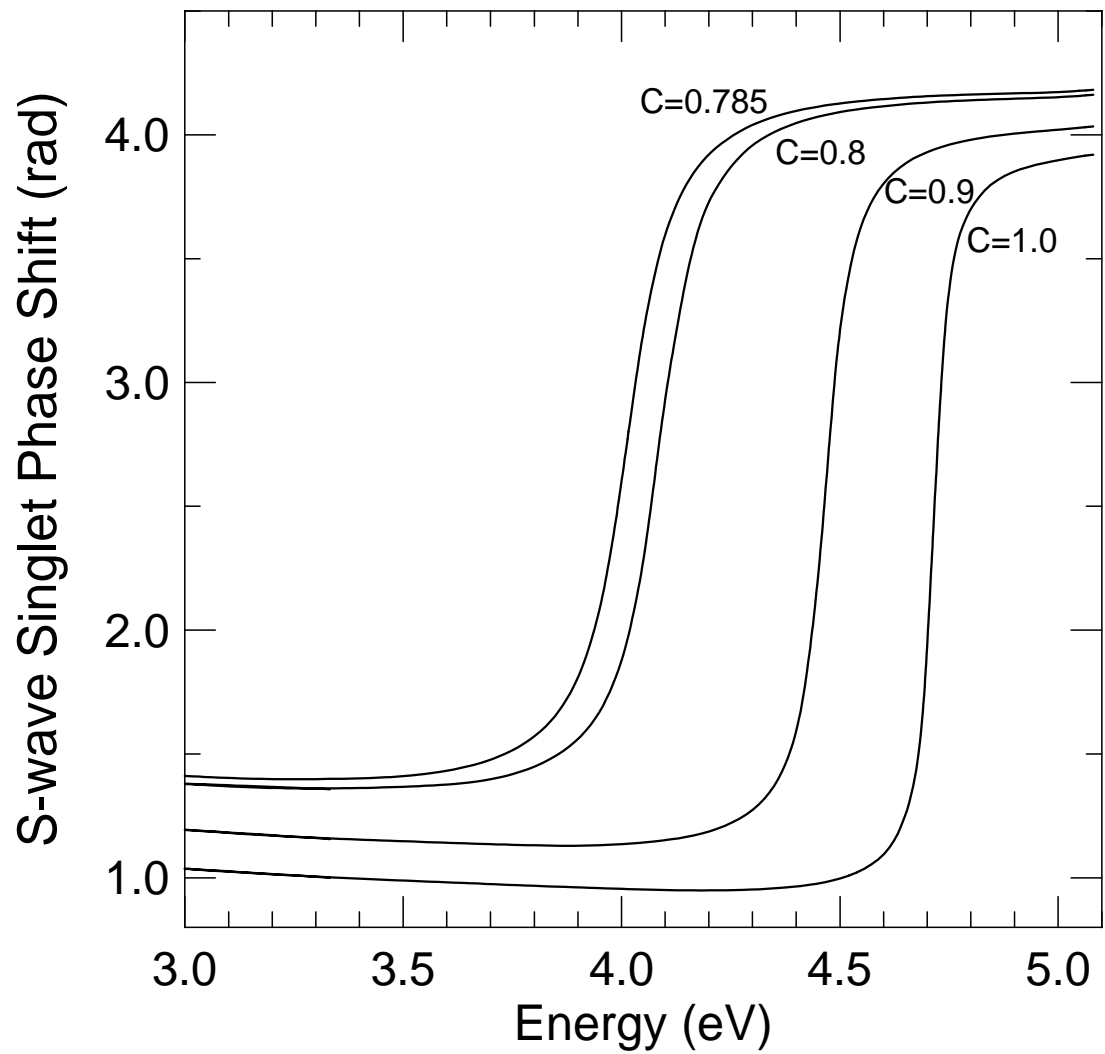
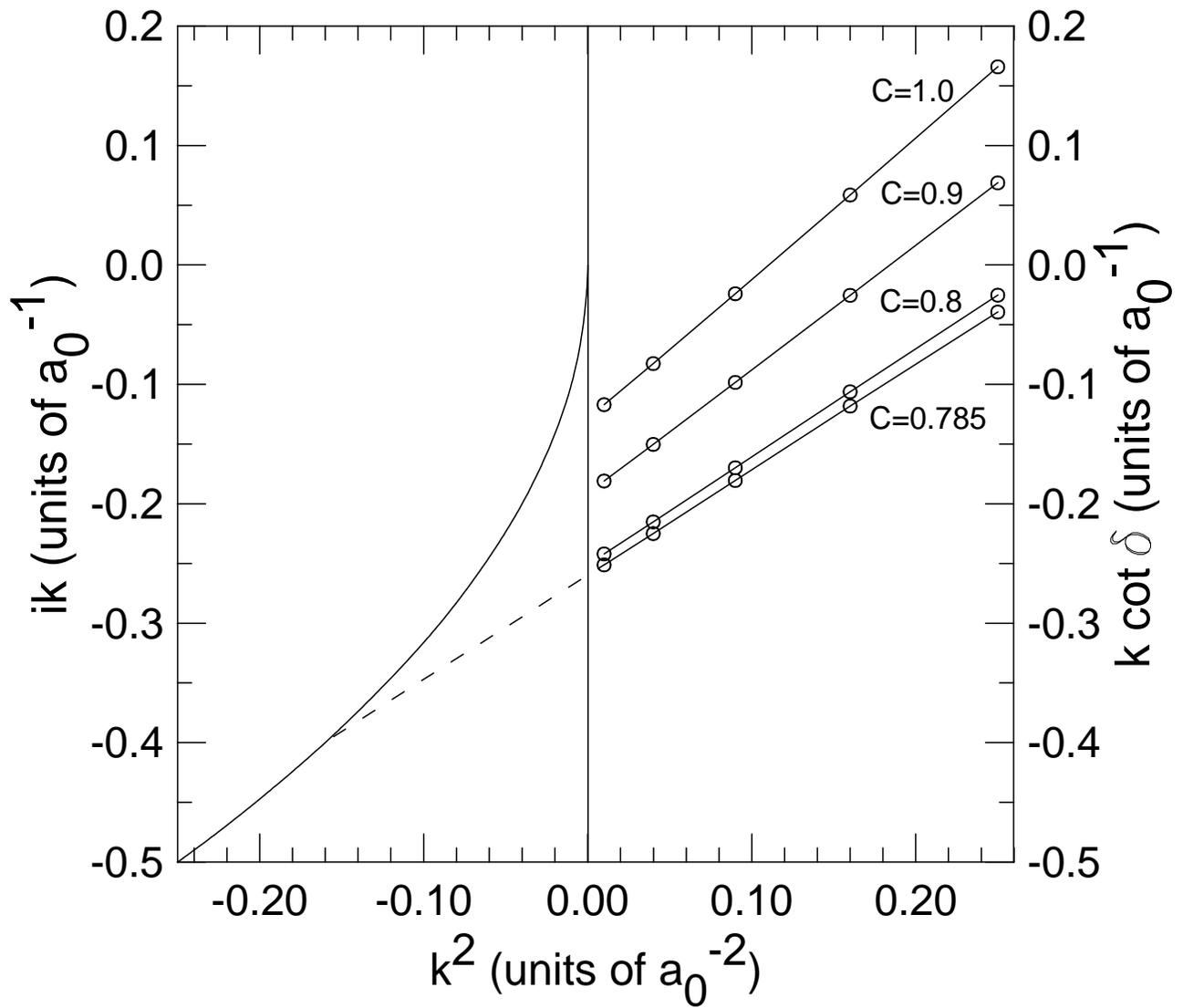
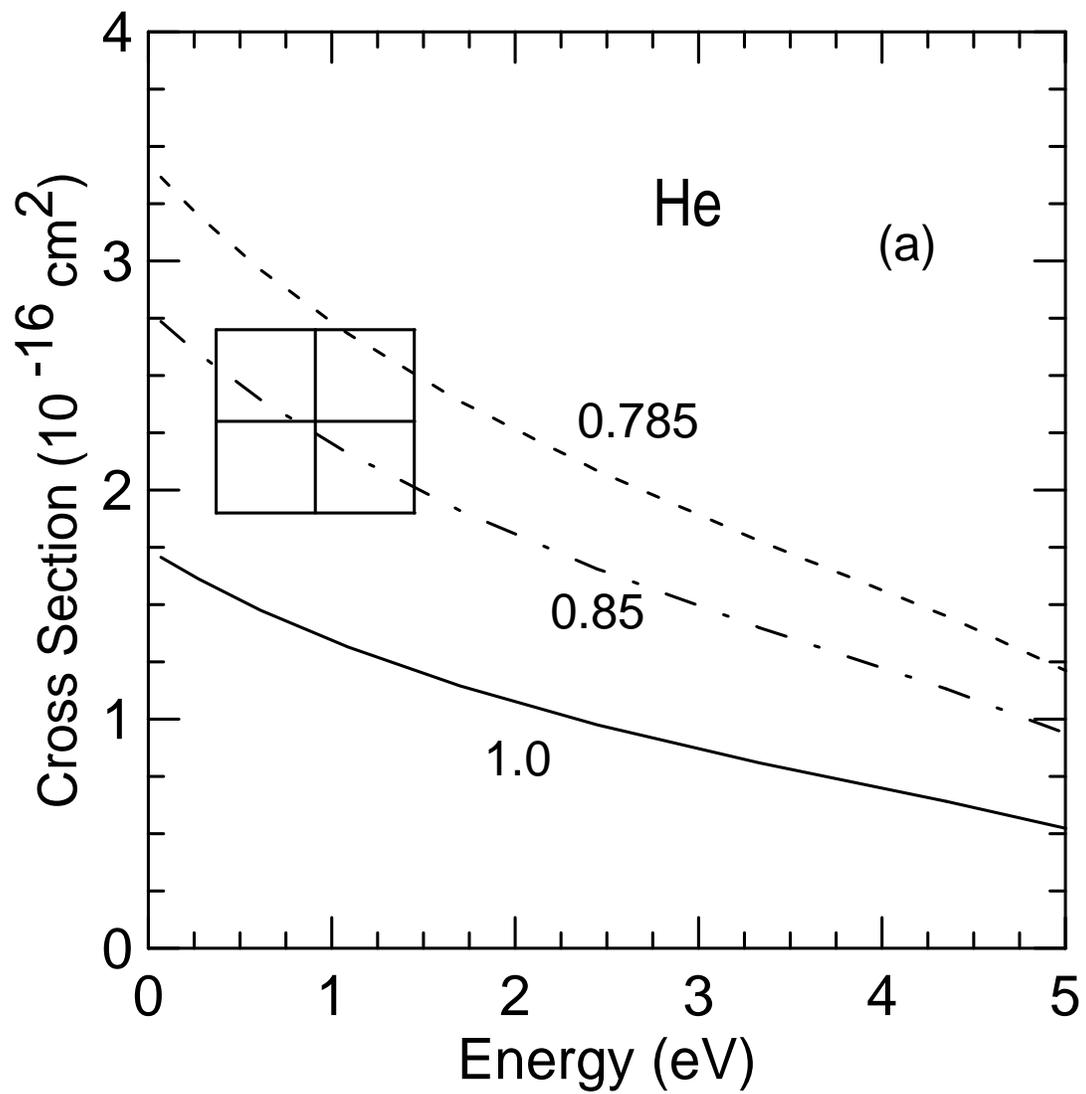
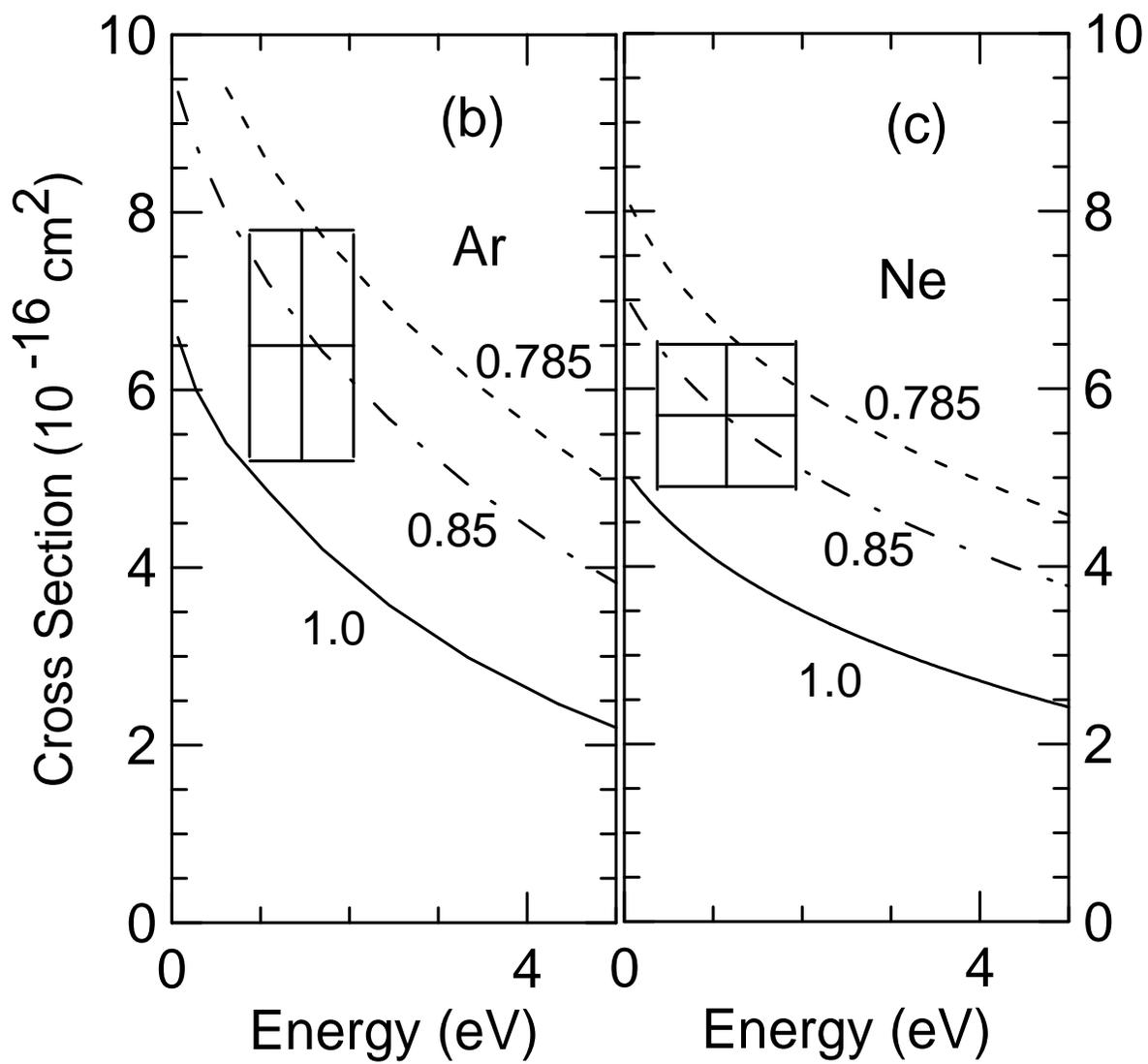
\end{document}